\documentclass[prl,twocolumn,floatfix,superscriptaddress]{revtex4}
\usepackage{amssymb}
\usepackage{amsmath}
\usepackage{graphicx}

\begin{document}

\title{Phase diagram and  critical exponents  of a dissipative Ising spin chain\\ in a transverse magnetic field}
\author{Philipp Werner}
\affiliation{Institut f{\"u}r theoretische Physik, ETH H{\"o}nggerberg, CH-8093 Z{\"u}rich, Switzerland}
\author{Klaus V\"olker}
\affiliation{Department of Physics, University of Toronto, Ontario, Canada M5S 1A7}
\author{Matthias Troyer}
\affiliation{Institut f{\"u}r theoretische Physik, ETH H{\"o}nggerberg, CH-8093 Z{\"u}rich, Switzerland}
\affiliation{Computational Laboratory, ETH Z\"urich, CH-8092 Z{\"u}rich, Switzerland}
\author{Sudip Chakravarty}
\affiliation{Department of Physics and Astronomy, University of California Los Angeles, Los Angeles, California 90095-1547, USA}
\date{\today}

\begin{abstract}
We consider a one-dimensional Ising model in a transverse magnetic field coupled to a dissipative heat bath. The phase diagram  and the critical exponents are determined from extensive Monte Carlo simulations. It is shown that the character of the quantum phase transition is radically altered from the corresponding non-dissipative model and the double-well coupled to a dissipative heat bath with linear friction. Spatial couplings and the dissipative dynamics combine to form a new quantum criticality which is independent of dissipation strength.
\end{abstract}

\maketitle

Dissipation in a quantum system is an important statistical mechanical problem having ramifications in systems as diverse as a single spin coupled to an environment \cite{Nature} to limitations of quantum computation \cite{Privman}. 
Of particular importance are the connections among decoherence, noise, dissipation, and the amount of coarse graining necessary for classical predictability \cite{MGM}. Dissipation can arise when we focus on some distinguished variables of a system, which are coupled to the ignored environment variables that are integrated out in a closed universe. Since dynamics and thermodynamics are intimately intertwined in quantum mechanics, dissipation plays a very important role. 

Although a single two-state system coupled to dissipation has been extensively discussed in the literature \cite{Leggett}, and its practical implications are abundant, the case of infinitely many spatially coupled two-state systems has not been discussed.  We shall show that spatial couplings and the dissipative dynamics combine to determine a new class of quantum critical points, which is different from the dissipative phase transition of the single two-state system \cite{Chakravarty} and the corresponding non-dissipative model. The nature of the quantum criticality does not depend on the value of the dissipation strength. This finding is in striking contrast to predictions for an array of resistively shunted Josephson junctions \cite{XY,Sumanta}.

The Hamiltonian for the Ising spin chain in a transverse field and coupled to a heat bath reads
\begin{eqnarray}
H&=&-J\sum_{i=1}^{N_x}\sigma^{z}_i\sigma^{z}_{i+1}-\Delta\sum_{i=1}^{N_x}\sigma^{x}_i\nonumber\\
&&+\sum_{i,k}\left\{C_k(a^\dagger_{i,k}+a_{i,k})\sigma^{z}_i+\omega_{i,k}a^\dagger_{i,k}a_{i,k}\right\},
\end{eqnarray}
where $\sigma^{x}$ and $\sigma^{z}$ are the Pauli matrices. To each site, $i$, we have coupled a set of independent bosons $\{a_{i,k}\}$ (destruction operators), of frequency  $\omega_{i,k}$, representing the environment.  The coupling strength, $C_k$, to the coordinate of the $k$th oscillator is  chosen such that the spectral function $J(\omega)=4\pi\sum_k {C_k^2}\delta(\omega-\omega_{i,k})=2\pi\alpha\omega$, for $\omega$ less than a cutoff frequency $\omega_c$, but vanishes otherwise. This spectral density defines linear dissipation whose strength is determined by the parameter $\alpha$ \cite{Caldeira}.

The bath degrees of freedom can be integrated out in a manner identical to the treatment of the spin-boson problem \cite{Luther}. The only difference is to recognize that the transfer matrix of the two-dimensional classical Ising model is the one-dimensional Ising model in a transverse field (non-dissipative, of course) \cite{Pfeuty}. The partition function defined on a $(1+1)$-dimensional lattice indexed by $i$ and $\tau$ is
\begin{equation}
Z=Z_{0}\sum_{\{\mu_{i,\tau}=\pm 1\}}e^{-S[\{\mu_{i,\tau}\}]},
\end{equation}
where $Z_{0}$ is the free boson partition function and 
\begin{eqnarray}
S&=&-K\sum_{i=1}^{N_x}\sum_{\tau=1}^{N_\tau}\mu_{i,\tau}\mu_{i+1,\tau}-\Gamma\sum_{i=1}^{N_x}\sum_{\tau=1}^{N_\tau}\mu_{i,\tau}\mu_{i,\tau+1}\nonumber\\
&&-\frac{\alpha}{2}\sum_{i=1}^{N_x}\sum_{\tau<\tau'}
\Big(\frac{\pi}{N_\tau}\Big)^2\frac{\mu_{i,\tau}\mu_{i,\tau'}}{\sin^{2}(\frac{\pi}{N_\tau}|\tau-\tau'|)} .
\label{1.10}
\end{eqnarray}
The classical Ising variables are defined by $\{\mu_{i,\tau}=\pm 1\}$, and the total number of lattice sites in the spatial and imaginary time directions are $N_{x}$ and $N_{\tau}$, respectively. Periodic boundary conditions are applied both in the space and imaginary time directions. 
The long-ranged term containing the dimensionless variable $\alpha$ describes the Ohmic dissipation \cite{Sandvik}. Here $K= \tau_{c} J$ and $e^{-2\Gamma}=\tanh (\tau_{c}\Delta)$, where $\tau_{c}$ is the lattice spacing in the imaginary time direction \cite{footnote}.  The universal scaling properties should not depend explicitly on $\tau_{c}$, and we can set it to unity without loss of generality. 
We shall investigate the critical behavior of this partition function as a function of the parameters $K$, $\Gamma$ and $\alpha$.

We use Monte Carlo simulations to study the action in Eq.~(\ref{1.10}), employing a variant of the Swendsen-Wang cluster algorithm \cite{Swendsen&Wang} developed by Luijten and Bl\"ote \cite{Luijten&Bloete}. This cluster algorithm greatly reduces  autocorrelation times and allows us to simulate lattices of more than $10^7$ spins. 
First, assuming a continuous phase transition, we attempt the scaling procedure of Ref. \cite{Rieger}.
\begin{figure}[t]
\centering
\includegraphics [angle=-90, width= 8.5cm] {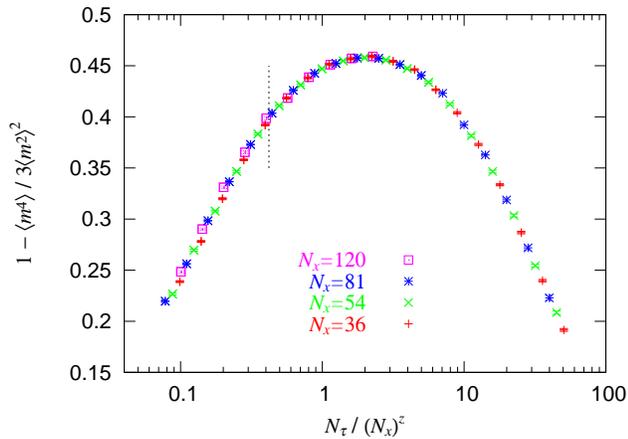}
\caption{Data collapse of Binder cumulants at the critical point for $\alpha=0.6$ and $z=2$. Points to the left of the dashed line were not considered since finite size effects become important in this region.}
\label{fig:collapse}
\end{figure} 
For each value of $\alpha$ we determine the critical coupling $K_c$ and the dynamical critical exponent $z$ self-consistently by looking for a data collapse of the Binder cumulant ratio $B=1-{\langle m^4\rangle}/({3\langle m^2\rangle ^2})$. Near the critical point $B$ scales as
\begin{equation}
B(N_x, N_\tau) = \phi_B\Big(\frac{N_{x}}{\xi}, \frac{N_\tau}{(N_x)^z}\Big). \label{3.1}
\end{equation}
In particular, at the critical point, where the correlation length $\xi$ diverges, the Binder cumulants collapse onto a universal function of $N_\tau/(N_x)^z$. 
The strategy is thus to plot $B$ as a function of $N_\tau$ for different values of $N_x$. 
At the critical coupling $K_c$ the maximum value of $B(N_x, N_\tau)$ is independent of $N_x$, and the critical exponent $z$ can be found from the optimal collapse of these data sets onto the universal curve $\phi_B(0,N_\tau/N_x^z)$, as shown in Fig. \ref{fig:collapse}. We use re-weighting techniques \cite{Ferrenberg&Swendsen} to vary the value of $K$ continuously and omit small system sizes where finite size corrections become important. The excellent data collapse validates the scaling {\em Ansatz} in Eq. (\ref{3.1}).

\begin{figure}[t]
\centering
\includegraphics [angle=-90,width=8.5cm] {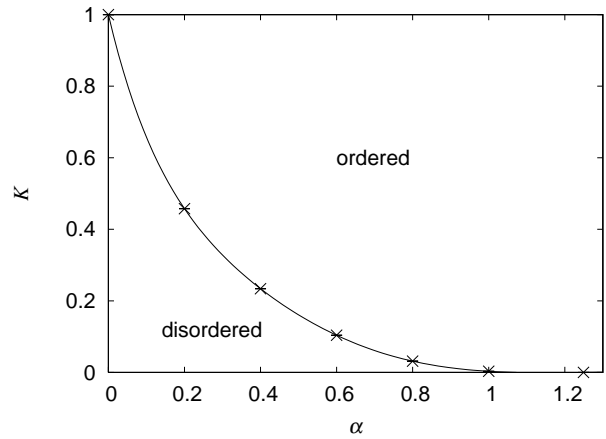}
\caption{Phase diagram of the dissipative quantum Ising chain in the space of nearest neighbor coupling $K$ and dissipation strength $\alpha$ for $\Gamma=0.136$. The phase boundary appears to approach the $K=0$ axis with zero slope, well approximated by an essential singularity.}
\label{fig:phasediagram}
\end{figure}

At first, we fix $\Gamma=-\frac{1}{2}\ln[\tanh(1)]=0.136$ and calculate the phase diagram as a function of $\alpha$ and  $K$. It is shown in Fig.~\ref{fig:phasediagram} and the corresponding values of $K_c$ and $z$ are listed in Tab.~\ref{table:exponents}. The critical point $\alpha^{*}$ at $K=0$ was computed using the method in Ref.~\cite{Luijten&Messingfeld} to be $\alpha^{*}=1.2492(4)$. The phase boundary $K_{c}(\alpha)$ seems to approach the $K=0$ axis with zero slope which points to an essential singularity.  This can be understood from a simple  mean-field argument, since the divergence of the susceptibility $\chi(\alpha\to \alpha^{*}_{-})$ at $K=0$ is exceedingly strong. The direction of approach  to the critical point in the fugacity-$\alpha$ plane does determine the precise form \cite{Chakravarty}, but the divergence has an essential singularity as $\alpha\to\alpha^{*}_{-}$ ($\chi(\alpha) \sim e^{A/\sqrt{\alpha^*-\alpha}}$). Therefore, for finite but small $K_{c}$, the phase boundary of the coupled system is  given by $2\chi(\alpha)K_{c}=1$, hence the result shown in Fig.~\ref{fig:phasediagram}.

We next consider the order parameter correlation function $C(x,\tau)$, which  in an infinite system scales as
\begin{eqnarray}
C(x,\tau)&=& \langle \sigma_{x,\tau}\sigma_{0,0}\rangle-\langle \sigma_{x,\tau}\rangle\langle \sigma_{0,0}\rangle \nonumber \\
& \sim & x^{-(z+\eta -1)}g(x/\xi,\tau/x^{z}).
\label{3}
\end{eqnarray}
The spatial correlation length $\xi$ diverges at the critical point and the characteristic frequency scale vanishes as $\xi^{-z}$, as long as the dynamical critical exponent $z$ is a finite and well defined quantity.  
At the critical coupling $K_c$, Eq.~(\ref{3}) simplifies to
\begin{eqnarray}
C(x,\tau) \sim x^{-(z+\eta -1)}\tilde{g}(\tau/x^z)=\tau^{-(z+\eta -1)/z}\hat{g}(x^z/\tau).
\label{4}
\end{eqnarray}
If we set $\tau=0$ and $K=K_c$, the spin-spin correlation function in the spatial direction should therefore decay asymptotically as $C(x,0) \sim x^{-(z+\eta-1)}$.
\begin{figure}[t]
\centering
\includegraphics [angle=-90, width= 8.5cm] {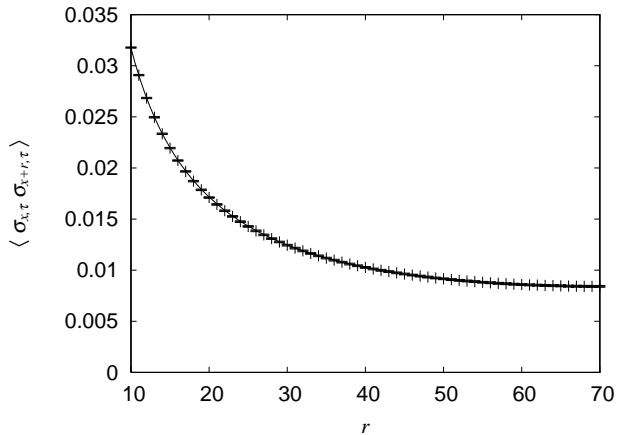}
\caption{Decay of the spin-spin correlation $\langle \sigma_{x,\tau}\sigma_{x+r,\tau}\rangle$ at the critical point for $\alpha=0.6$. The size of the lattice is $N_x=140$ and $N_\tau=10^5$. The curve shows the best fit to the data points in the interval $10\le r \le 70$.}
\label{fig:corr}
\end{figure} 
Observing that $N_\tau\gtrsim 5N_x^z$ is sufficient to avoid finite size effects in the $\tau$-directions 
we chose $N_x=140$ and $N_\tau=10^5$ and show the result in Figure \ref{fig:corr}. 
Fits were performed to the functional form $(x^{-k}+(N_x-x)^{-k})$ in the region $x_{\text{min}} \le r \le N_x/2=70$. The optimal value for $k=z+\eta-1$ decreases from 1.003 for $x_{\text{min}}=5$ to $0.99$ for $x_{\text{min}}=20$ while the error-bars increase in size. We estimate $z+\eta=2.00(1)$.
\begin{figure}[htb]
\centering
\includegraphics [angle=-90, width=8.5cm] {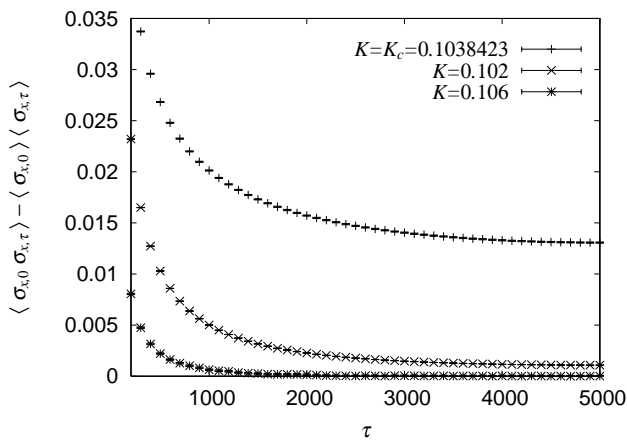}
\caption{Spin-spin correlation function in imaginary time for $\alpha=0.6$, $N_x=500$ and $N_\tau=10^4$. The three curves show (from top to bottom) the decay at the critical point, and in the disordered and ordered phase respectively.}
\label{fig:timecorr}
\end{figure} 

The imaginary time correlation function, shown in Fig.~\ref{fig:timecorr} falls to zero exponentially on the ordered side  and $\langle \sigma_{x,\tau}\sigma_{0,0}\rangle$ approaches $\langle \sigma_{x,\tau}\rangle^2$.  On the disordered side, the correlation function cannot decay faster than $\tau^{-2}$ \cite{Griffith} and does not allow one to define a correlation time and a dynamical critical exponent $z$ except  in the sense of a generalized ``Josephson length'' \cite{Chakravarty2}. Taking into account the observed large corrections to scaling, our results at the critical point $K=K_c$, shown  in Fig.~\ref{fig:timecorr} for $N_x=500$ and $N_\tau=10^4$ are roughly consistent with Eq. (\ref{4}) and the previously determined values of $z$ and $\eta$. On the other hand, it was shown that for the corresponding $(0+1)$-dimensional problem $C(\tau)\sim 1/\ln \tau$ at criticality \cite{Bhattacharjee}.  Because of smaller corrections to scaling, we used the spatial dependence of the correlation function for  the quantitative analysis.
\begin{table}[htb]
\caption{Critical coupling and the critical exponents $z$ and $\nu$ for different values of the dissipation strength $\alpha$ and the coupling $\Gamma$.} 
\centering
\begin{tabular}{lllllllll}
  \hline
  \hline
  $\Gamma$ && $\alpha$ && $K_c$ && $z$ && $\nu$\\
  \hline
  0.136 && 0 && 1 && 1 && 1\\
  0.136 && 0.2 && 0.457699(5) && 1.97(3) && 0.637(7)\\
  0.136 && 0.4 && 0.23396(1) &&  && \\
  0.136 && 0.6 && 0.103842(2) && 2.00(2) && 0.639(3)\\
  0.136 && 0.8 && 0.03142(1) &&  && \\
  0.136 && 1.0 && 0.0031045(1) && 1.97(4) && 0.624(13)\\
  0.136 && 1.2492(4) && 0 && -- && --\\
  \hline
  1.153 && 0.6 && 0.0011169(2) && 1.98(4) && 0.623(9)\\
  \hline
  \hline
\end{tabular}
\label{table:exponents}
\end{table}
\begin{figure}[t]
\centering
\includegraphics [angle=-90, width= 8.5 cm] {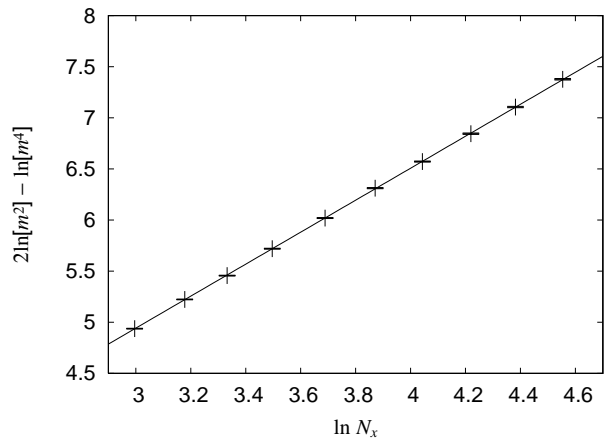}
\caption{Plot of $2\ln [m^2]-\ln [m^4]$  as a function of $\ln N_x$ at $\alpha=0.6$ and fixing $N_\tau=5(N_x)^z$). The quantity $[m^n]$ is defined in Eq. (\ref{3.8}) and according to Eq. (\ref{3.7})  the slope of the fitted line gives $1/\nu$.}
\label{fig:nu}
\end{figure}

We finally determine the critical exponent $\nu$. By choosing $N_\tau=5(N_x)^z$ we again reduce the two-parameter scaling to a one-parameter one and can employ the same procedure as for the classical Ising model \cite{Ferrenberg&Landau}. From the scaling ansatz
$\langle m^n \rangle = N_x^{-n\beta/\nu}\phi_n((K-K_c)N_x^{1/\nu})$ it follows that
\begin{equation}
2\ln\left.\frac{\partial \langle m^2\rangle}{\partial K}\right|_{K_c} - \ln\left.\frac{\partial \langle m^4\rangle}{\partial K}\right|_{K_c} = \text{const}+\frac{1}{\nu}\ln N_x.\label{3.7} 
\end{equation}
The derivatives of $\langle m^n\rangle$ with respect to $K$ can be calculated from the correlation function
\begin{equation}
\left.\frac{\partial \langle m^n\rangle}{\partial K}\right|_{K_c} = {\langle m^n \Sigma_{x}\rangle}_{K_c} - {\langle m^n \rangle}_{K_c}{\langle \Sigma_{x} \rangle}_{K_c} =:[m^n],\label{3.8} 
\end{equation}
where $\Sigma_x$ denotes the quantity $\Sigma_{x}=\sum_{x,\tau}\sigma_{x,\tau}\sigma_{x+1,\tau}$.
Results for $\nu$, obtained from fits such as shown in Fig.~\ref{fig:nu} are listed in Tab.~\ref{table:exponents}. The errors due to the uncertainty in $z$ were estimated from simulations with system sizes $N_\tau=5(N_x)^z$, using both $z=1.98$ and $z=2$. Within error bars there is no dependence of the critical exponents on the dissipation strength.

To test if our choice of the third parameter in the action in Eq.~(\ref{1.10}) -- the nearest neighbor coupling $\Gamma$  -- affects the values of the critical exponents, we repeated the simulations for $\Gamma=-\frac{1}{2}\ln[\tanh(0.1)]$ at $\alpha=0.6$ and found consistency with the previous results. Note that when the Hamiltonian of the  Ising model in a transverse field contains the coupling to the heat bath, $\Gamma$ is an independent parameter.  We finally checked that our numerical results do not depend on the aspect ratio by increasing $N_\tau/(N_x)^z$ from $5$ to $20$ and used two different random number generators
\cite{LaggedFibonacci, Mersenne}
to rule out the only remaining possible sources of systematic errors.
\begin{table}
\centering
\caption{Comparison of exponents obtained by our simulations and the values predicted by the $\epsilon$-expansion. The values of $\nu$ and $z$ are averages of the various estimates listed in Tab.~\ref{table:exponents}. The sum $z+\eta$ was determined for $\alpha=0.6$ only.}
\begin{tabular}{llll}
  \hline
  \hline
  & simulation&   one loop & two loop   \\
  \hline
  $\nu$ & $0.638\pm0.003$  & 0.583  \cite{Pankov} & 0.633 + $O(\epsilon^3)$ \cite{Sachdev} \\
  $\eta$ & $0.015\pm0.020$ & 0 & 0.020 + $O(\epsilon^3)$ \cite{Pankov} \\
  $z$ & $1.985\pm0.015$ & 2 & 1.980 + $O(\epsilon^3)$ \cite{Pankov} \\
  $z+\eta\hspace{2mm}$ & $2.00\pm0.01\hspace{7mm}$ & 2 & 2 \cite{Pankov} \\
  \hline
  \hline
\end{tabular}
\label{tab:compare}
\end{table}

We now compare these result with the predictions from a dissipative $n$-component  $\phi^{4}$ field theory \cite{Pankov,Sachdev} in an expansion in $\epsilon=2-d$ (the one dimensional Ising chain corresponds to $n=1$ and $\epsilon=1$), which is based on their result that the dissipation strength does not renormalize to all orders in $\epsilon$. The agreement with the simulation shown in Table \ref{tab:compare} is very good, but may deteriorate with higher order terms, as even the  $\epsilon$-expansion of the standard $\phi^{4}$ theory is asymptotic in nature and reliable exponents cannot be obtained without a Borel-Pad{\'e} analysis \cite{Zinn}. 

We believe that for the first time we have been able to carry out  a precise simulation of an extended dissipative quantum system and have made progress towards understanding of quantum criticality in the presence of dissipation. An intriguing problem that can be analyzed by the present method  is the criticality of a $(0+1)$-dimensional transition embedded in the higher dimensional system discussed in the context of an array of resistively shunted Josephson junctions   \cite{XY,Sumanta} and the quantum theory of the smectic metal state in stripe phases \cite{Emery}. It is remarkable, however, that a floating phase arises naturally in the problem of a dissipative Josephson junction array, where the states of $(0+1)$-dimensional power law phases can slide past one another.  

S. C. has been supported by the NSF under Grant No. DMR-0411931. M.T. acknowledges support by the Swiss National Science Foundation. The calculations have been performed on the Asgard Beowulf cluster at ETH Z\"urich, using the ALPS library \cite{ALPS}. We thank F.~Alet, J.~Fr\"ohlich, Ch.~Helm,  S.~Kivelson, C.~Nayak, S.~Sachdev and S.~Tewari for many interesting discussions. S. C. and M. T. would like to thank the Aspen Center for Physics, where a part of the work was carried out.


\begin{thebibliography}{99}
\bibitem{footnote} Alternative discretizations in the Trotter-Suzuki formula can yield different results, although they all agree in the limit $\tau_{c}\to 0$. For example, another expression for $\Gamma$ is $-\frac{1}{2}\ln [\tau_{c}\Delta]$. Since, we are only interested in the universal critical properties, these differences are of no consequence.
\end{thebibliography}
\end{document}